\documentclass[prb,article,superscriptaddress,preprint]{revtex4-1}

\usepackage{amssymb}
\usepackage{amsfonts}
\usepackage{amsmath}
\usepackage{graphicx}
\usepackage{dcolumn}
\usepackage{bm}
\usepackage{xcolor}
\usepackage{float}
\usepackage{textgreek}
\usepackage{upgreek}
\usepackage[colorlinks]{hyperref}
\usepackage[normalem]{ulem}
\usepackage{soul}
\hypersetup{
                pdfstartview={FitH},
                linkcolor=blue,
                citecolor=blue,
                filecolor=blue,
                urlcolor=blue
}

\begin{document}


\title{Synthesis of epitaxial monolayer Janus SPtSe}

\author{Roberto Sant}
\affiliation{Univ. Grenoble Alpes, CNRS, Institut N\'{e}el, 38042 Grenoble, France}
\affiliation{ESRF, the European Synchrotron, 38043 Grenoble, France}

\author{Maxime Gay}
\affiliation{Univ. Grenoble Alpes, CEA, LETI, 38054 Grenoble, France}

\author{Alain Marty}
\affiliation{Univ. Grenoble Alpes, CEA, IRIG/DePhy/SpinTec, 38054 Grenoble, France}

\author{Simone Lisi}
\affiliation{Univ. Grenoble Alpes, CNRS, Institut N\'{e}el, 38042 Grenoble, France}

\author{Rania Harrabi}
\affiliation{Univ. Grenoble Alpes, CEA, IRIG/DePhy/MEM/NRS, 38054 Grenoble, France}

\author{C\'eline Vergnaud}
\affiliation{Univ. Grenoble Alpes, CEA, IRIG/DePhy/SpinTec, 38054 Grenoble, France}

\author{Minh Tuan Dau}
\affiliation{Univ. Grenoble Alpes, CEA, IRIG/DePhy/SpinTec, 38054 Grenoble, France}

\author{Xiaorong Weng}
\affiliation{Univ. Grenoble Alpes, CEA, IRIG/DePhy/MEM/NRS, 38054 Grenoble, France}

\author{Johann Coraux}
\affiliation{Univ. Grenoble Alpes, CNRS, Institut N\'{e}el, 38042 Grenoble, France}

\author{Nicolas Gauthier}
\affiliation{Univ. Grenoble Alpes, CEA, LETI, 38054 Grenoble, France}

\author{Olivier Renault}
\affiliation{Univ. Grenoble Alpes, CEA, LETI, 38054 Grenoble, France}

\author{Gilles Renaud}
\email{gilles.renaud@cea.fr}
\affiliation{Univ. Grenoble Alpes, CEA, IRIG/DePhy/MEM/NRS, 38054 Grenoble, France}

\author{Matthieu Jamet}
\email{matthieu.jamet@cea.fr}
\affiliation{Univ. Grenoble Alpes, CEA, IRIG/DePhy/SpinTec, 38054 Grenoble, France}


\maketitle



\textbf{Janus single-layer transition metal dichalcogenides, in which the two chalcogen layers have a different chemical nature, push chemical composition control beyond what is usually achievable with van der Waals heterostructures. Here we report such a novel Janus compound, SPtSe, which is predicted to exhibit strong Rashba spin-orbit coupling. We synthetized it by conversion of a single-layer of PtSe$_2$ on Pt(111) via sulphurization under H$_2$S atmosphere. Our \textit{in situ} and \textit{operando} structural analysis with grazing incidence synchrotron X-ray diffraction reveals the process by which the Janus alloy forms. The crystalline long-range order of the as-grown PtSe$_2$ monolayer is first lost due to thermal annealing. A subsequent recrystallization in presence of a source of sulphur yields a highly ordered SPtSe alloy, which is iso-structural to the pristine PtSe$_2$. The chemical composition is resolved, layer-by-layer, using angle-resolved X-ray photoelectron spectroscopy, demonstrating that Se-by-S substitution occurs selectively in the topmost chalcogen layer.}



Most electronic properties of crystals are inherited from their symmetries. For monolayer transition metal dichalcogenides (TMDCs) in the 1H phase like MX$_2$ (M=Mo,W and X=S, Se), the mirror symmetry with respect to the transition metal atoms plane leads to zero internal out-of-plane electric field and suppresses any Rashba spin-orbit coupling (SOC). The same happens for monolayer TMDCs in the 1T phase like PtSe$_2$ because of crystal inversion symmetry. In both phases, the application of a vertical electric field lowers the symmetry and can induce spin-splitting of the electronic bands due to the Rashba SOC\cite{riley,yuan}. The resulting spin texture is at the origin of unique spintronics phenomena\cite{edelstein}. 
An alternative route consists in breaking the out-of-plane mirror symmetry (in 1H phases) or the crystal inversion symmetry (in 1T phases) by substituting one of the two chalcogen layers with a layer of another chalcogen species. The resulting ternary compounds, so-called Janus after the biface Roman god, are predicted to exhibit strong Rashba SOC, possibly altered by mechanical stress and external electric fields\cite{cheng, hu}. The internal electric field in Janus TMDCs may be exploited also in piezoelectric devices\cite{dong} and p-n junctions\cite{palsgaard}. Anionic (chalcogen) substitution reactions have been used to tune the composition of TMDCs, with the main purpose of engineering the bandgap\cite{ma,feng,taghinejad}. However, few reports addressed so far the selective substitution of the chalcogen atoms in a single layer to prepare ordered Janus compounds\cite{lu, zhang2017}. Single layer (SL) 1T PtSe$_2$ is an indirect bandgap semiconductor\cite{wang} with a high spin-orbit interaction. It exhibits interesting magnetic properties, such as helical spin texture\cite{yao}, hidden spin polarization connected to spin-layer locking\cite{yao}, defect-induced magnetic ordering\cite{avsar} and magnetic anisotropy \cite{zhang}. The first epitaxial synthesis of SL PtSe$_2$ was performed by direct selenization of a Pt(111) crystal surface\cite{wang}. Since then, other growth approaches have been reported, including chemical vapour deposition\cite{besenbacher}, molecular beam epitaxy\cite{yan}, plasma assisted selenization\cite{su} and thermally assisted conversion of predeposited metal layers\cite{yim}.\\
Here, we synthesized epitaxial SL PtSe$_2$ by selenization of a Pt(111) single-crystal, and we developed a method for the conversion of the as-grown material into the Janus SPtSe by sulphurization of PtSe$_2$ using H$_2$S gas. The whole process has been monitored \textit{in situ} under ultrahigh vacuum (UHV) or under H$_2$S partial pressure of 10$^{-4}$ mbar by synchrotron grazing incidence X-ray diffraction (GIXRD). These experiments: (i) confirmed a long-range ordered coincidence lattice superstructure stemming from the mismatch between PtSe$_2$ and Pt(111)\cite{wang}; (ii) revealed a perfect epitaxial alignment of PtSe$_2$ along the main Pt(111) crystallographic directions but with sizeable lattice distortions in both the PtSe$_2$ and the topmost Pt(111) layers, in contrast to what would occur if the epitaxy was of van der Waals (vdW) type and (iii) proved the structural \textit{pinning} of the 2D layer to the substrate; the coincidence lattice remaining unaltered after the sulphurization of the pristine PtSe$_2$. Finally and most importantly, we demonstrated that sulphurization under carefully controlled temperature conditions leads to a layer-selective substitution, and hence to an ordered Janus material, where S atoms replace Se ones only in the top layer, as demonstrated by angle resolved X-ray photoemission spectroscopy (AR-XPS). Longer exposure to H$_2$S at higher temperature further allows to form a PtS$_2$ SL isostructural to the pristine PtSe$_2$ one and pinned to the Pt(111) substrate by substituting all the Se atoms with S atoms.\\

\textbf{PtSe$_2$ growth by selenization of Pt(111) and structural analysis}\\
Before we address the formation and structure of the Janus alloy, we discuss about its parent compound, the pristine SL PtSe$_2$. SL PtSe$_2$ was grown in UHV by a two-step process based on the direct selenization of a Pt(111) crystal surface (see Methods for details) as described by Wang \textit{et al.}\cite{wang}. Firstly, several nm of amorphous selenium were deposited on top of Pt(111) until the reflection high-energy electron diffraction (RHEED) pattern disappeared. Secondly, the sample was annealed from room temperature (RT) to 370$^\circ$C. During the annealing, the platinum atoms bond to selenium at the interface forming a PtSe$_2$ monolayer, whereas the excess selenium desorbs from the surface.\\

Fig. \ref{fig:xrd_asgrown}a shows an 80$^\circ$ in-plane sector of the reciprocal space mapped with GIXRD after the selenization. This map displays the successive Brillouin zones of Pt(111) (orange hexagons) having sharp Pt(111) peaks at their centres, with additional sharp peaks corresponding exactly to a (4$\times$4) superstructure relative to the Pt(111) lattice. In Fig. \ref{fig:xrd_asgrown}b, two high-resolution linear scans measure the scattered intensity along the radial direction $h$ (highlighted by the red frame in Fig. \ref{fig:xrd_asgrown}a) before (blue) and after (orange) the selenization. The Pt peaks are pointed out by blue filled circles. The PtSe$_2$ peaks, labelled with orange filled circles, are characterized by a periodic reciprocal lattice vector which is three quarters the Pt(111) surface one, consistent with previous reports\cite{wang}. Additional peaks, marked by orange diamonds in between the Pt and PtSe$_2$ peaks are perfectly aligned and equi-spaced. This diffraction pattern is the signature of a superlattice which is the coincidence between (3$\times$3) PtSe$_2$ and (4$\times$4) Pt(111) surface cells, and has a superperiodicity of 11.1 \AA, in agreement with Ref.\citenum{wang}.
The reciprocal space is indexed using this superstructure unit cell, so that the Pt(111) and PtSe$_2$ Bragg peaks are at $h$ and $k$ surface reciprocal lattice units ($s.r.l.u.$) multiple of 4 and 3 respectively. We deduced a PtSe$_2$ surface lattice constant of ($3.700\pm 0.002$) \AA, slightly smaller (by 0.7 \%) than the relaxed bulk value of 3.724 \AA, and a very good epitaxial alignment between the PtSe$_2$ monolayer and the Pt substrate. 
From the analysis of the width of PtSe$_2$ Bragg peaks in transverse scans (see Fig. \ref{fig:xrd_asgrown}c and Supplementary Note 2), we deduced a finite domain size of 240 \AA\ and a negligibe in-plane mosaicity. From the analysis of radial scans along the ($h00$) and ($hh0$) crystallographic directions (see Supplementary Note 2), we obtained a very similar domain size of 220 \AA\ and an estimation of the relative inhomogeneous strain of $\Delta a/a\approx 3\times10^{-6}$. 
The global compressive strain of 0.7 \% and the perfect angular alignment are signatures of a strong interaction between PtSe$_2$ and Pt(111). This hypothesis is supported by the remarkably strong intensity of the superstructure peaks, similar or exceeding those of PtSe$_2$ peaks nearby ($e.g.$ compare peaks at $h=7$ and $h=6$). It indicates sizeable distortion in the two surface lattices.
In Fig. \ref{fig:xrd_asgrown}d (other rod scans are shown in Supplementary Fig. 1), the intensity profiles recorded along the $l$ direction perpendicularly to the surface and in correspondence of the superstructure peaks positions exhibit broad peaks and dips. The perpendicular coordinate $l$ is expressed in (surface) reciprocal lattice units ($s.r.l.u.$) of the hexagonal Pt unit cell with basal plane parallel to the Pt(111) surface, and of parameter $c=6.797$ \AA\ equal to the thickness of three Pt(111) planes in the bulk. Notably, most superstructure peaks (except those of PtSe$_2$ with $h$ and $k$ multiple of 3), are located close to integer values of $l$ (see $e.g.$ ($50l$) rod in Fig. \ref{fig:xrd_asgrown}d), which are  Bragg conditions for Pt(111). As already discussed in the past for other surface systems\cite{croset}, this indicates that the interlayer periodicity in the substrate contributes substantially to the superstructure rods shapes and that lattice distortions are present, not only in the PtSe$_2$ SL, but also in the few topmost Pt layers. Moreover, from the superstructure rods, we can confirm the monolayer character of PtSe$_2$ (see Supplementary Note 3).\\
Obviously, our PtSe$_2$/Pt(111) system cannot be described using the van der Waals epitaxy picture where the 2D overlayer grows almost strain-free due to very weak interaction with its substrate. It rather implies a strong coupling between PtSe$_2$ and the Pt(111) surface. Such a strong interaction has important consequences in the structural constraints inherited by the material upon sulphurization. In particular, as shown below, it helps keeping the integrity of the PtSe$_2$ crystal during annealing and after the sulphurization process.\\

\textbf{Stability of the as-grown PtSe$_2$ under annealing} \\
\textit{In situ} measurements during annealing in UHV have been performed to determine the temperature window inside which the epitaxial PtSe$_2$ SL is stable. Fig. \ref{fig:xrd_sulphurisation}a displays ($h00$) radial scans of the scattered X-ray intensity containing the second order PtSe$_2$ reflection (at $h=6$), two superstructure peaks (at $h=5$ and $h=7$) and two Pt(111) peaks (at $h=4$ and $h=8$). The superstructure and PtSe$_2$ peak intensities progressively decrease with time when the annealing temperature exceeds the growth temperature of 370$^\circ$C, until they vanish completely (spectra labeled a-d in Fig.~\ref{fig:xrd_sulphurisation}a). We deduce that above the growth temperature the annealing not only lifts the superstructure order (\textit{i.e.} the PtSe$_2$ and Pt lattice constants are not in a 3-to-4 ratio anymore), but also degrades the PtSe$_2$ structure. Surprisingly, cooling down to 300$^\circ$C or lower leads to the recrystallization of the PtSe$_2$ layer and restores the original superstructure, as evident in the diffraction patterns (compare spectra labeled d-g and RT in Fig.~\ref{fig:xrd_sulphurisation}a). The most plausible scenario accounting for these observations is that the Se-Pt bonds break at annealing temperatures above 370$^\circ$C, and yield a disordered, possibly highly mobile, phase of selenium atoms onto the substrate surface, as suggested by the background increase in the scans taken at intermediate temperatures (not shown) compared to the RT scan. The degradation of PtSe$_2$ and the concomitant release of Se atoms appear to be progressive, occurring over a time range of several tens of minutes. The domain size is apparently not affected since the width of the in-plane peaks is unchanged. This suggests point defects, for instance chalcogen vacancies, as the dominant type of disorder in the system. Cooling at RT restores the pristine structure, \textit{i.e.} free Se atoms are consumed for the recrystallization of PtSe$_2$, \textit{e.g.} by filling vacancies. This also implies that the Se desorption rate is marginal at the explored temperatures.\\

\textbf{Sulphurization of PtSe$_2$ in H$_2$S atmosphere}\\ 
So far, we have discovered that a reversible path can be followed from dissolution to recrystallization of PtSe$_2$ upon cycles of annealing and cooling. At this point, we wondered whether, after degradation of the PtSe$_2$, one could follow a new cooling path in the presence of a sulphur precursor, leading to an ordered chalcogen alloyed SL TMDC, \textit{via} substitution of some Se atoms by S ones, thanks to the higher stability of Pt-S bonds compared to Pt-Se ones. To test this proposal, we supplied sulphur in the form of gaseous H$_2$S,\cite{gronborg,sanders2,sanders} which is known to decompose on noble metal surfaces. 

We prepared a first sample called SPtSe by pre-annealing it in UHV a few tens of degrees above the growth temperature (between 370$^\circ$C and 400$^\circ$C), in order to create Se vacancies in the topmost chalcogen layer, which will be eventually filled with S atoms. Then, a 10$^{-4}$ mbar H$_2$S partial pressure was introduced in the UHV chamber and maintained for five hours at a slightly lower temperature of 350$^{\circ}$C. We monitored the sulphurization process by GIXRD to track possible structural changes due to the chalcogen substitution in the PtSe$_2$ lattice, \textit{e.g.} strain, loss of the superstructure order, surface disorder, new phases with different lattice parameters, or additional/suppressed peaks corresponding to different crystal symmetries. In Fig. \ref{fig:xrd_sulphurisation}b, the radial scans of as-grown PtSe$_2$ and sulphurized PtSe$_2$ are compared. After H$_2$S treatment, the intensity loss (-58\%) of the (600) PtSe$_2$ Bragg peak and the broader tail in the low $h$ values of the Pt(400) crystal truncation rod (CTR) peak suggest an increased surface disorder\cite{robinson}. Except for that, the spectra are remarkably similar to those of pristine PtSe$_2$/Pt(111), the (3$\times$3)-on-(4$\times$4) coincidence being exactly maintained. The chemical composition of this SPtSe sample was probed by XPS. The Se \textit{3s}/S \textit{2s}, Se \textit{3p}/S \textit{2p} and Se \textit{3d} core levels spectra are shown in Fig. \ref{fig:xps_all} and compared to those of a reference sample of pristine PtSe$_2$ prepared by selenization. We clearly see almost equal contributions from S and Se demonstrating the partial sulphurization of this sample.\\

In order to achieve full sulphurization, we prepared a second sample called PtS$_2$ by exposing PtSe$_2$ to H$_2$S partial pressure for a longer time (9 h in total) after a pre-annealing in UHV at a higher temperature (up to 460$^\circ$C). The XPS spectra of this PtS$_2$ sample in Fig. \ref{fig:xps_all} show a clear predominant sulphur content with little traces of selenium. Noteworthy, the very intense Se Auger lines at 174 and 178.5 eV visible for PtSe$_2$ are particularly weak for this sample. The sulphurization was also monitored \textit{in operando} by GIXRD. Similarly to Fig. \ref{fig:xrd_sulphurisation}b, Fig. \ref{fig:xrd_sulphurisation}c shows two radial scans along the $h$ direction taken before (black) and after (red) the sulphurization process. Here, the peak intensity at $h=6$ (previously PtSe$_2$ reflection) is severely damped to about one tenth of the pristine intensity and becomes asymmetric, the intensity at the Pt CTR tails (on the low-$h$ sides of the $h=4$ and $h=8$ peaks) is higher and all the peaks are wider, all suggesting disorder in the surface layer. The PtSe$_2$ overlayer is very likely degraded due to the higher temperature used and to the longer exposure to H$_2$S. However, remarkably, a residual ordered superstructure is still preserved, which can be hardly associated to any pure PtSe$_2$ or SPtSe alloy  because of the lack of selenium determined by XPS. No other diffraction signals were found which could be attributed to a relaxed platinum disulfide (PtS$_2$) or other phases. It is worth noting that for PtS$_2$ to be in a 4:3 ratio with Pt(111), conditions required by the superstructure, it should stretch its lattice parameter by as much as 4.6 \%. The expected Bragg peak position in Fig. \ref{fig:xrd_sulphurisation}c is indicated by a black arrow at $h=6.28$. Therein, the only new detectable signatures are some peak shoulders, that we ascribe to either sulphide phases of slightly shorter lattice parameter than PtSe$_2$ coexisting with the  superstructure or the presence of defects yielding a diffuse scattering contribution. A plausible interpretation of our observations is that the H$_2$S supply at elevated temperature has allowed for a replacement of all Se atoms by S atoms. Similarly, no substantial modifications of the pristine diffraction pattern were observed in the small map of Fig. \ref{fig:xrd_sulphurisation}d. The only significant changes are the clear elongation of most of the superstructure peaks toward the (440) PtSe$_2$ main reflection. These elongations might arise from specific defects that would need a very extensive investigation, which is beyond the scope of the present study. Some scattering rods produced by the superstructure were also measured on both sulfurized samples (Fig. \ref{fig:xrd_sulphurisation}e). Except for their different intensity and faster decrease with $l$, they show qualitatively exactly the same features as the original PtSe$_2$ ones. This confirms that, even for the fully sulfurized sample, the original PtSe$_2$ structure is preserved. 

To summarize, the in-plane diffraction pattern and the out-of-plane rods of the as-grown PtSe$_2$ remain almost unchanged upon the sulphurization procedures while S atoms substitute some or all of Se atoms (Fig. \ref{fig:xrd_sulphurisation}b-e). It shows that both the SPtSe and PtS$_2$ alloys are isostructural to the pristine PtSe$_2$ (although with some disorder that we attribute to the creation of defects). Hence, all along the recrystallisation, the memory of the initial PtSe$_2$ pinning on its substrate is maintained, and the substitution of the chalcogen atoms occurs without significant modification of their lateral position with respect to the substrate atoms. Both the strong interaction between Pt and PtSe$_2$ and the fortunate 4 to 3 coincidence of the in-plane lattice parameters are believed to lead to this $pinning$. This conclusion is a further evidence that the epitaxy in this system, rather than being of vdW type, is governed by more covalent interface interactions leading to distortions in both the overlayer and the substrate. These interactions probably explain why the PtSe$_2$ crystal retains its crystal integrity upon sulphurization. In the following, we demonstrate that the sulphurization process of the SPtSe sample follows a scenario similar to that depicted in Fig. \ref{fig:xrd_sulphurisation}f.\\

\textbf{Evidence for a Janus alloy}\\
To probe the chemical composition of the SPtSe layer and estimate the vertical distribution of S and Se atoms, we performed AR-XPS measurements, as illustrated in Fig. \ref{fig:xps_janus}a. Prior to the measurements, the samples were systematically annealed in vacuum in order to remove any trace of H$_2$S that might be adsorbed at the surface after growth. In AR-XPS, the photo-emission intensity varies with the depth of excitation of the photoelectrons which is proportional to $\sin \theta$, $\theta$ being the take-off angle between the surface and the spectrometer. The lower the take-off angle is, the smaller the probed depth and the higher the surface sensitivity are. Exploiting this effect allows determining whether sulphur substitutes selenium in the TMDC and whether the substitution occurs in the top or in the bottom layer, or in both. Figs.~\ref{fig:xps_janus}b-d show the Se \textit{3p} and S \textit{2p} core-level (CL) spectra after subtraction of a Shirley background, fitted by two spin-orbit doublets for three different take-off angles  $\theta$=45$^\circ$, 25$^\circ$ and 10$^\circ$. The spin-orbit splitting of the Se \textit{3p} and S \textit{2p} CLs are fixed at 5.75 and 1.18 eV, respectively\cite{vergnaud,cadot}. We find that the S contribution increases when $\theta$ decreases, while the opposite trend is observed for Se. The ratio between the S and Se contributions to the spectra decreases as a function of the probed depth, from 1.54 at $\theta=10^{\circ}$ (short depth) to 1.15 at $\theta=45^{\circ}$ (long depth), as shown in Fig. \ref{fig:xps_janus}e. This result clearly shows that the H$_2$S treatment induces a Se-by-S substitution in the PtSe$_2$ lattice preferentially in the top layer. It suggests also that the kinetic barriers to the substitution differ for the two chalcogen layers. 

We then performed a quantitative analysis of the Se and S CL intensities derived from the peak fitting of Fig. \ref{fig:xps_janus}b-d and from the spectra of Fig. \ref{fig:xps_all} to determine the chalcogen vertical distribution in SPtSe by AR-XPS. This analysis faces two main challenges. Firstly, elastic scattering along the path taken by the photoelectrons towards the detector\cite{tilinin} needs to be taken into account in a rigorous way.  This effect is most prominent at grazing take-off angles as used here, and results in significant changes in the measured intensities emitted from buried layers. The correction procedure for elastic scattering is complex because it varies with the geometry of the experiment and the details of the sample composition. Secondly, when a sample consists of chemically different layers, the exact element-specific photoelectron inelastic mean free path (IMFP) must be taken into account for each of these layers.

\begin{table}[ht!!]
\begin{center}
\begin{tabular}{|c|c|c|c|c|}
\hline 
\textbf{Photoelectron/layer} & Pt5p & Se3p & S2p & Se3d \\ 
\hline 
Kinetic energy (eV) & 1450 & 1300 & 1300 & 1450 \\ 
\hline 
IMFP (\AA) & 15.7 & 26.2 & 29.5 & 28.5 \\ 
\hline 
Asymmetry parameter & 1.75 & 1.30 & 1.16 & 1.08 \\ 
\hline 
Intensity at 10$^{\circ}$ & 626.2 & 240.3 & 188.6 & 147.8 \\ 
\hline 
Intensity at 25$^{\circ}$ & 2702.6 & 924.9 & 648.9 & 541.1 \\ 
\hline 
Intensity at 45$^{\circ}$ & 6076.1 & 1561.5 & 813.0 & 850.1 \\ 
\hline 
\end{tabular}
\end{center}
\caption{Kinetic energy, IMFP, asymmetry parameter, and intensity values at three take-off angles for the quantitative analysis of the AR-XPS intensities.}
\label{Table1}
\end{table}

For this work, we used the QUASES-ARXPS software developped by Tougaard \textit{et al.} (http://quases.com/products/quases-arxps/). In this approach, elastic scattering is accounted for by a relatively simple analytical expression of the depth-distribution function which describes the measured angular-dependent intensity\cite{nefedov1,nefedov2}. To determine quantitatively the Se and S concentrations in each of the chalcogen layers, we have considered as physical constraint and input the thickness of the SPtSe overlayer, and let the concentrations be derived from the angle-dependent core-level intensity curve. The SPtSe thickness was calculated as follows: we used the attenuation law of the Pt substrate signal, considering the Pt $4f$ CL at $\approx$71.5 eV before and after the formation of SPtSe. We found a thickness of 3.7 \AA\ (the details can be found in Supplementary Note 4). This value is close to the PtSe$_2$ SL thickness (2.53 \AA\ in Ref.\citenum{wang}) and confirms the monolayer character of SPtSe, consistent with the analysis of diffraction data (Supplementary Note 3). The other important inputs are related to the IMFPs and the asymmetry parameter which takes into account the geometry of the measurement at each take-off angle. The IMFP is both element-specific and also depends on the photoelectron kinetic energy; it is determined from the Tanuma-Powell-Penn formula using the QUASES-IMFP routine. The asymmetry parameter is derived from Ref.\citenum{band} and accounts for the experiment geometry through the electron orbital symmetry and the angle between X-rays and the sample surface. All the input parameters are summarized in Table \ref{Table1}.


Finally, we determined the Se concentration in the bottom chalcogen layer from the Se \textit{3d} angle-dependent intensities, after fitting the Se \textit{3d}-Pt \textit{5p$_{3/2}$} CL spectra corrected for elastic scattering. This calculation showed that the Se signal originates at 98.5 \% from the bottom chalcogen layer, with only negligible contributions both from the Pt layer and the topmost chalcogen one. We next considered the angle-dependent intensities of both Se \textit{3p} and S \textit{2p} CL spectra of Fig. \ref{fig:xps_janus}b-d, and found equal Se and S total concentrations in the sample. We thus conclude that, within the estimated 15 \%-20 \% uncertainty, the selenium is localized exclusively in the bottom layer, whereas all the Se atoms of the first layer have been substituted by S atoms. This result establishes that the sulphurization process leads to a highly ordered ternary compound in a Janus configuration isostructural to the pristine PtSe$_2$ monolayer: in the conditions we used, thermal energy is just enough for S atoms to selectively substitute the first Se layer as schematically depicted in Fig. \ref{fig:xrd_sulphurisation}f.\\
In comparison, we also performed \textit{ex situ} AR-XPS on the PtS$_2$ sample. Those measurements demonstrated that a quasi-complete substitution of selenium by sulphur has occurred in this case. A very low Se concentration (3.5 \%) is estimated at 45$^\circ$ take-off angle for the \textit{Se 3p} CL (for more grazing angles the estimate comes with a prevailing uncertainty due to the lower signal-to-noise ratio). In these harsher sulphurization conditions, thermal energy seems high enough, and the time of the process sufficiently long, for S atoms to diffuse through the PtSe$_2$ in order to reach and occupy selenium vacancies in the bottom Se layer. This entails overcoming the kinetic barriers that allow selective substitution of the first chalcogen layer to achieve the complete substitution of Se atoms by S atoms and the formation of a PtS$_2$ monolayer isostructural to the pristine PtSe$_2$ one.

\textbf{Conclusion}\\
In this work, we focused on the transformation of single layer PtSe$_2$ into an ordered Janus 2D SL SPtSe. We found that epitaxial PtSe$_2$ is strongly bonded to its Pt(111) substrate. This situation is at variance with vdW epitaxy, since here the PtSe$_2$ layer is strained and highly oriented. PtSe$_2$ was next converted into an epitaxial Janus SPtSe TMDC by exposure to H$_2$S at a temperature at which its structure is strongly altered. While cooling without H$_2$S leads to a recrystallisation towards the initial PtSe$_2$, the presence of the sulphur molecular precursor for well chosen temperature and exposure time allowed the transformation of the top Pt-Se bonds into Pt-S ones. The differing kinetic barriers for the Se-by-S  substitution process in the top and bottom chalcogen layers allowed recrystallization in a Janus configuration. This highly ordered 2D TMDC, which is not found in nature, exhibits crystal inversion symmetry breaking. This is expected to translate into new physical properties, \textit{e.g.} a spontaneous electric polarization and a piezoelectric character, as well as strong Rashba spin-orbit coupling.


\newpage

\section*{Methods}
PtSe$_2$ was prepared in two different UHV chambers hosted in Grenoble (France). First, Pt(111) single crystal surfaces were prepared in the INS2 UHV CVD-MBE-diffraction chamber at the CRG/IF-BM32 beamline\cite{cantelli} (\url{www.esrf.eu/UsersAndScience/Experiments/CRG/BM32}) at the European Synchrotron radiation Facility (ESRF, \url{www.esrf.eu}). Subsequently, selenium was deposited in a second MBE chamber located at the CEA Grenoble. Base pressures in the two systems are in the low 10$^{-10}$mbar range. The selenization process for the formation of PtSe$_2$ has been followed by \textit{in-situ} X-ray diffraction in the same INS2 UHV environment where the Pt(111) crystals were prepared. Samples were kept under UHV conditions (10$^{-8}$ mbar) during the transfers between the two chambers by means of a UHV home-made suitcase compatible with the two UHV setups. Finally, XPS measurements were performed in a PHI 5000 VersaProbe II photoelectron spectrometer at the Nanocharacterization Platform (PFNC) of the Minatec Campus in Grenoble (France).

\textbf{Samples preparation}\\
The preparation of the (111) surface of the single crystals (MaTech) consists of several cycles of Ar$^+$ ion bombardment (initially vigorous at 1.0kV and then milder at 0.8 kV) at $3\times10^{-6}$ mbar Ar partial pressure followed by annealing in UHV at 900$^\circ$C, until the surface shows a sharp and unreconstructed RHEED pattern. Amorphous selenium was deposited on the bare Pt(111) crystal surfaces at room temperature under a $10^{-6}$ mbar partial pressure of Se for two minutes. This corresponds to a several nm thick amorphous Se film. The selenization is achieved by annealing the selenium covered sample in UHV up to the optimized temperature of 370$^\circ$C inside the INS2 chamber. Temperatures were measured with pyrometers. The formation and optimization of the PtSe$_2$ structure and of the sulphurized structures were monitored in situ by GIXRD.

\textbf{INS2 Diffractomer}\\
The selenization process and final PtSe$_2$ structure were studied at INS2 by GIXRD. In INS2, the $z$-axis diffractometer is coupled with a bending magnet X-ray source (using a bending magnet) and the growth chamber, allowing structural characterization and growth at the same time. The beam energy was set at 19.8 keV and the incident angle at 0.18$^\circ$, slightly below the critical angle for total reflection. The beam size on the sample is 25(H)$\times$300(V) $\mu$m$^2$ whereas the vertical divergence is as low as 0.13 mrad. The studied samples are 8 mm in diameter and are vertically mounted. The diffracted intensity is acquired with a MaxiPix 2D hybrid pixel detector (1296$\times$256 pixels, 55$\times$55 $\mu$m$^2$ pixel size) developed at the ESRF and located at 0.7 m from the diffractometer homocenter. 

\textbf{GIXRD data acquisition and integration}\\
Measurements of the in-plane diffracted intensity in the reciprocal space plane have been performed by means of rocking scan centered on the PtSe$_2$/superstructure peaks\cite{drnec}. The analysis (2D data plotting, integration of the collected intensities) has been performed using \textit{PyRoD} (software specialized in the treatment of GIXRD  data taken with a 2D detector, under development at ESRF-BM32).

\textbf{Angle-resolved XPS}\\
The data were measured with a PHI 5000 VersaProbe II photoelectron spectrometer at the Nanocharacterization Platform (PFNC) of the Minatec Campus in Grenoble (France). We transferred the samples between the INS2 UHV growth chamber and the
XPS chamber under inert nitrogen atmosphere to limit the contamination from ambient air. The samples were inserted in ultra-high vacuum and excited by a 200 $\mu$m-diameter monochromatic Al-K$\alpha$ (h$\nu$=1486.6 eV) X-ray light source. The source-analyzer angle is 45$^\circ$ and remains unchanged during the AR-XPS data acquisition where the photoelectron take-off angle (sample-analyzer angle) varies. The spectral intensity also decreases drastically in the same time, imposing a higher acquisition time. The overall energy resolution (taking into account both the spectrometer and X-ray bandwidths) is 0.6 eV for core level spectra and we set the C \textit{1s} peak from the adventitious carbon to 284.8 eV for calibration of the binding energy scale. Data correction and peak deconvolution were performed with the CasaXPS software and we used the sensitivity factors provided by the instrument to perform the relative quantification of the elements.

\section*{Data Availability}
The data that support the findings of this study are available from the corresponding author (GR \& MJ) upon reasonable request.

\section*{Supplementary Information}
Supplementary Information comprises additional measurements (diffraction rod $l$-scans), the details about the peak width analysis in GIXRD, the estimation of the PtSe$_2$ thickness by GIXRD and the estimation of the SPtSe thickness by XPS.

\section*{Acknowledgements}
The authors acknowledge the financial support from the French ANR project MAGICVALLEY (ANR-18-CE24-0007), the 2DTransformers project under the OH-RISQUE program (ANR-14-OHRI-0004), the French state Equipex funds (ANR-11-EQPX-0010, ANR-10-LABX-51-01) (LABEX LANEF) in the framework of the \textit{Programme d'Investissements d'Avenir}. The authors also acknowledge the European Synchrotron Radiation Facility for provision of beam time on the BM32 beamline. Part of this work was performed at the Platform For Nanocharacterization of the CEA-Grenoble, MINATEC Campus, with support from the French \textit{Recherche Technologique de Base} programme. We thank Maurizio De Santis for providing the single crystal substrate used in the experiments.

\section*{Author Contributions}
G.R. and M.J. conceived and coordinated the project. Discussions in presence of G.R., M.J, O.R., J.C and R.S were regularly scheduled to interpret the results and elaborate the next strategies. R.S. and A.M., assisted by T.D and C.V, prepared the substrate crystals before the allocated beam times; R.S., A.M., S.L. and G.R. performed the experiments during the beam times at BM32 (PtSe$_2$ growth and sulfurization, GIXRD characterization); O.R designed the AR-XPS experiment and M.G performed the measurements and the data reduction; R.S., R.H and X.W., supervised by G.R., quantitatively analysed the diffraction data; N.G. and O.R. performed the quantitative analysis of the AR-XPS data. R.S., G.R., M.J., J.C. and O.R. cowrote the paper.

\section*{Competing interests}
The authors declare no competing interests.

\newpage

\section*{Figure captions}

Figue 1: \textbf{GIXRD characterization of PtSe$_2$2 grown by selenization of Pt(111). a,} 80$^{\circ}$ sector map of the reciprocal space parallel to the surface at $l=0$,  as acquired after the selenization; ($h, k, l$) are surface reciprocal lattice units expressed in terms of the coincidence supercell; the orange hexagonal frames delimit Pt(111) Brillouin zones; orange and blue circles highlight two in-plane peaks belonging respectively to PtSe$_2$ and the Pt substrate lattices. The rings of scattered intensity centered around the origin can be indexed as polycrystalline grains Pt reflections, very likely originated from damages areas of the single crystal surface. The inset on the right shows a highly-resolved diffraction pattern corresponding to the black-dashed-line-delimited sector in the extended map; the image has been acquired from a different sample without Pt polycrystalline grains. \textbf{b}, Highly-resolved in-plane radial scans measured along the high symmetry direction $h$ (highlighted in red in \textbf{a}) of the  Pt(111) substrate before (blue) and after (orange) the selenization; the $h$ axis is indexed using supercell reciprocal lattice units ($s.r.l.u.$); Blue and orange filled circles indicates Pt(111) and PtSe$_2$ reflections, whereas the remaining peaks are associated to superstructure peaks by orange diamonds. \textbf{c}, Plot and linear fit of the squared transversal width of the in-plane peaks measured by rocking scans as a function of the inverse squared modulus of the transfer vector $\bf{q}$. The intercept of the fit with the vertical axis provides an estimation of the (negligible) overlayer mosaic spread. The slope of the linear fit allows deducing the average PtSe$_2$ domain size. \textbf{d}, Scans along the out-of-plane direction $l$ perpendicular to the sample surface at the positions ($33l$), ($50l$) and ($70l$), relative to a  PtSe$_2$ Bragg peak and two superstructure peaks. The logarithm of the intensity is plotted.\\

\begin{figure}
   \centering
   \includegraphics[scale=0.85]{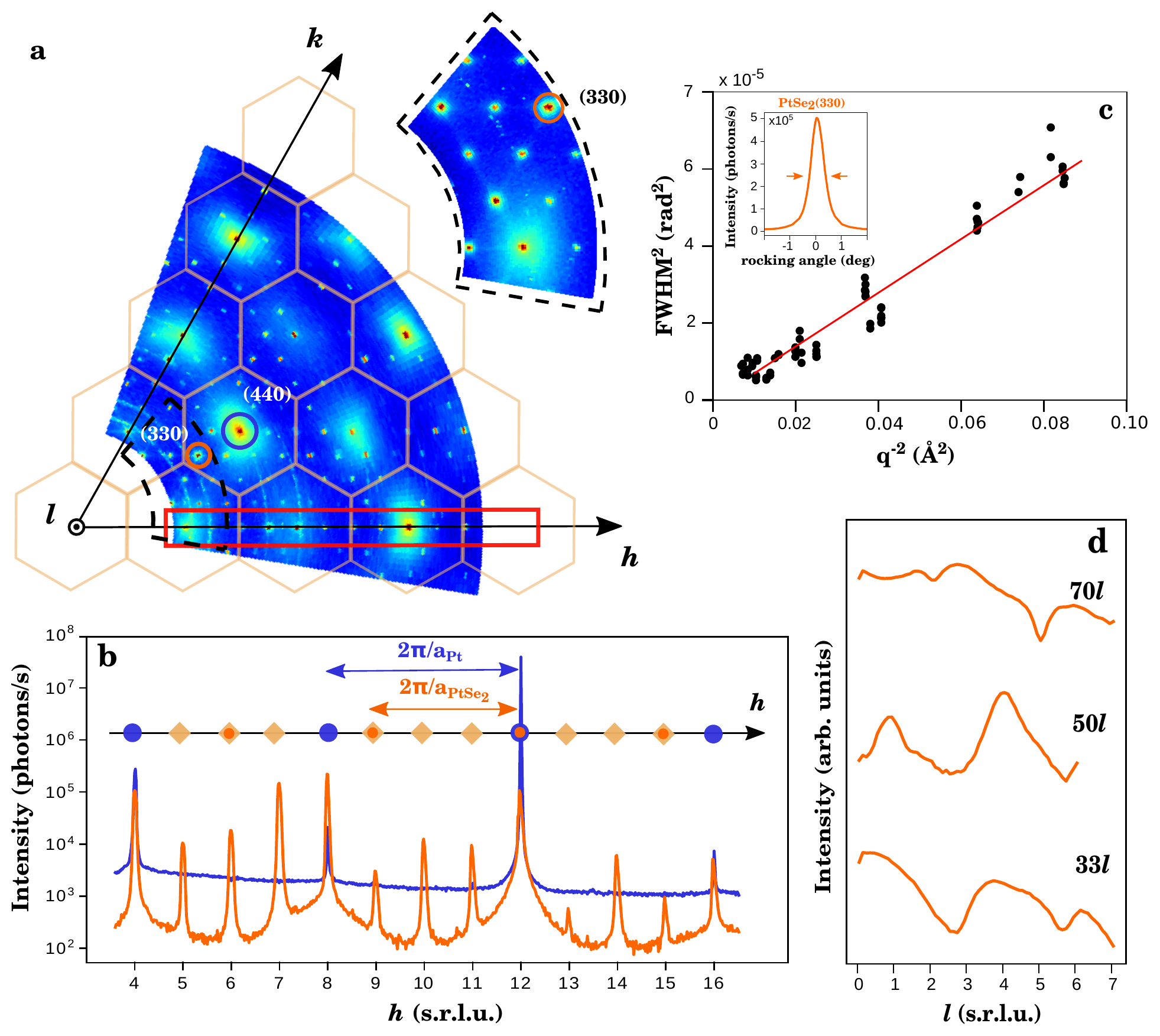}
    \caption{}
   \label{fig:xrd_asgrown}
\end{figure}

\begin{figure}
   \centering
    \includegraphics[scale=0.75]{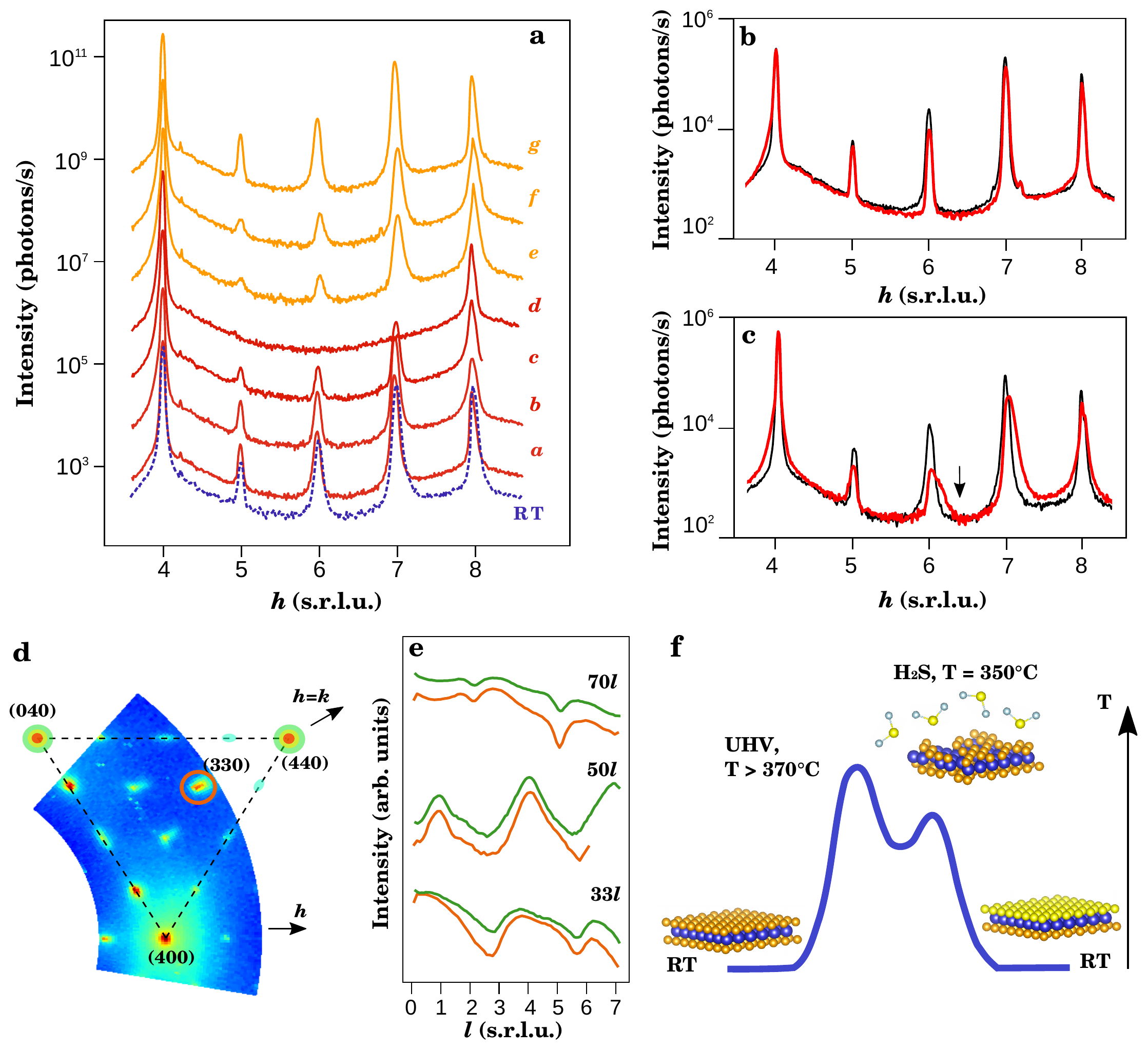}
    \caption{}
    \label{fig:xrd_sulphurisation}
\end{figure}

\begin{figure}[ht]
    \centering
    \includegraphics[scale=0.63]{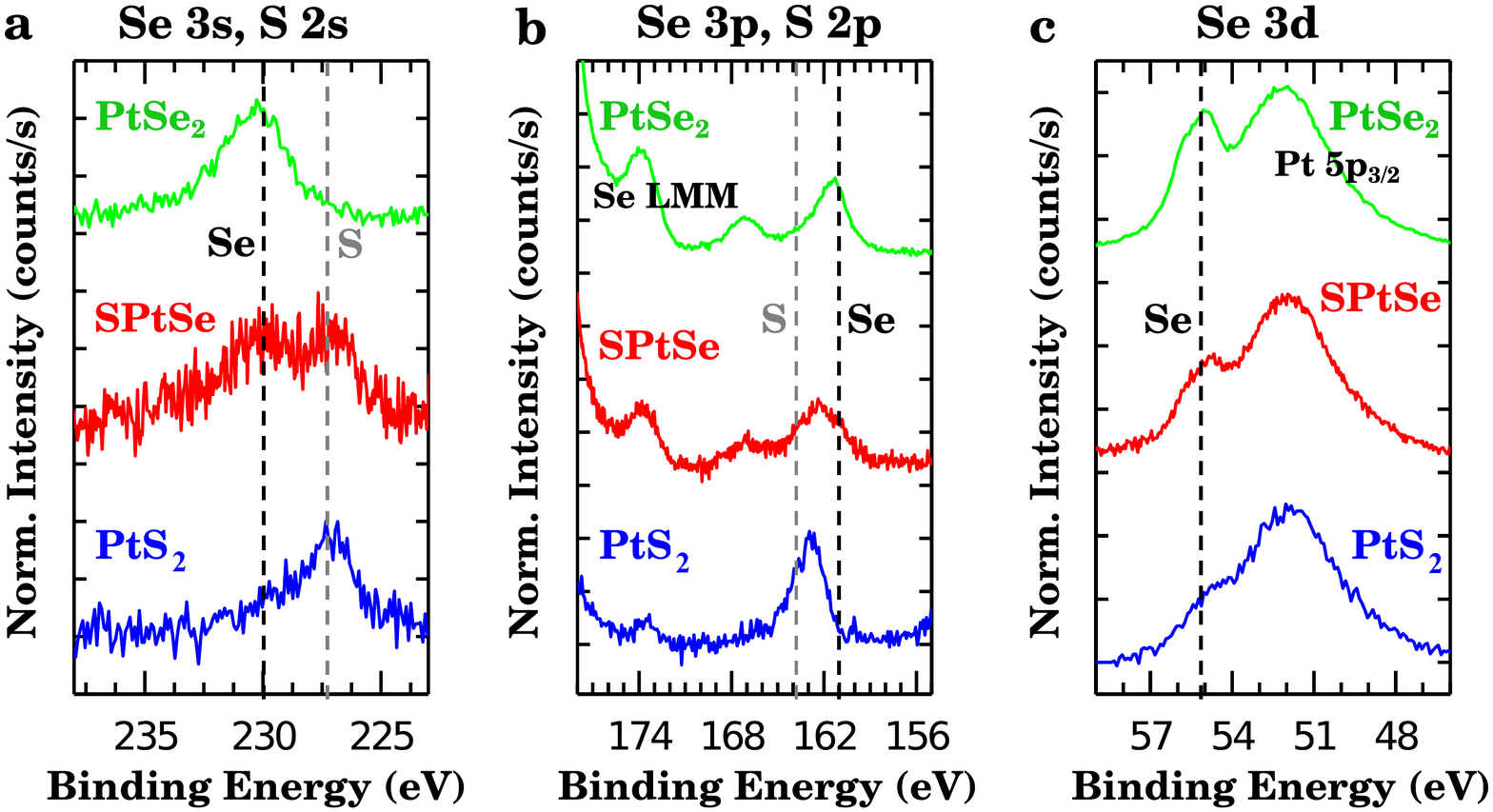}
    \caption{}
    \label{fig:xps_all}
\end{figure}

\begin{figure}[ht]
    \centering
    \includegraphics[scale=0.95]{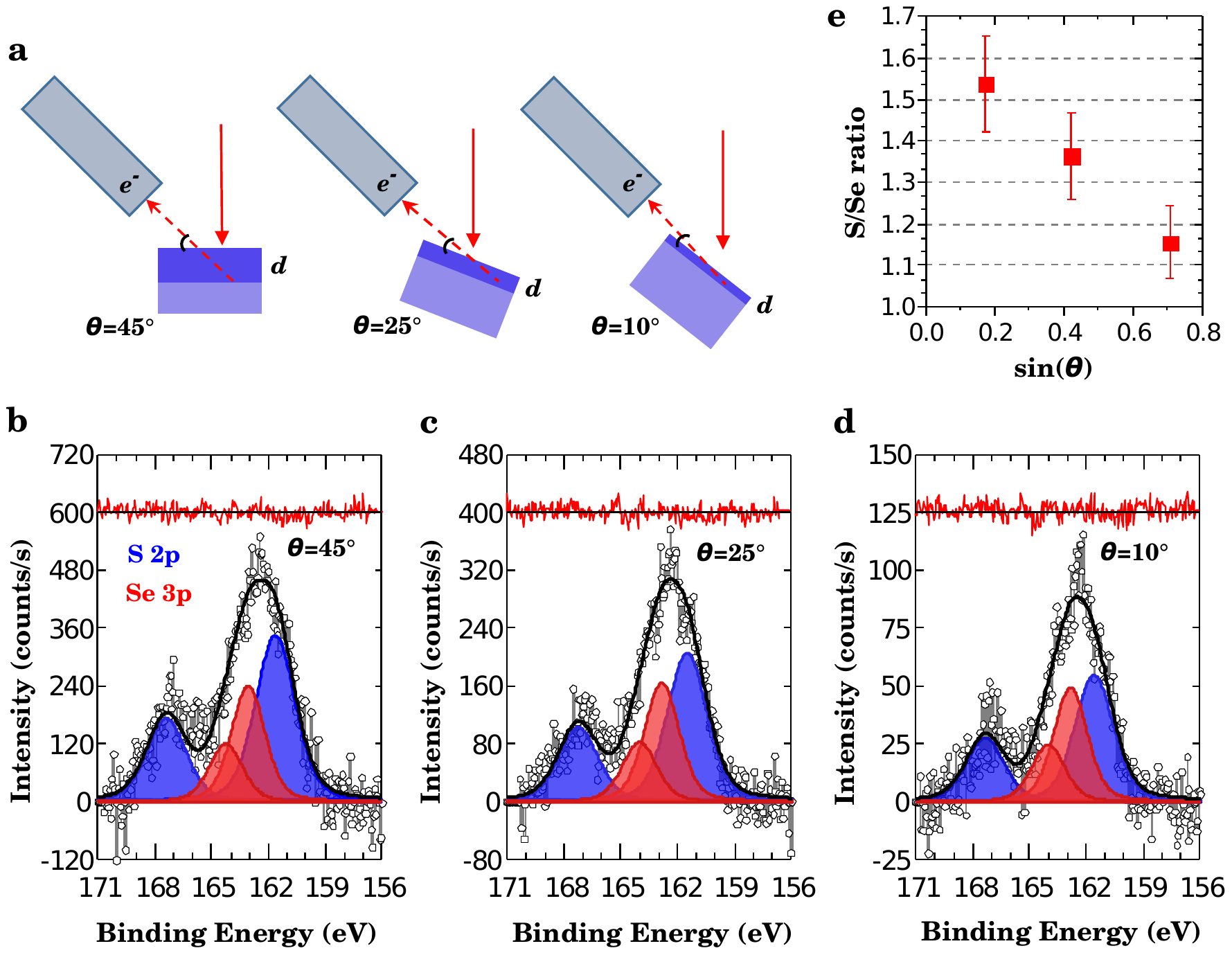}
    \caption{}
    \label{fig:xps_janus}
\end{figure}

Figure 2: \textbf{Annealing and sulphurization of PtSe$_2$ in H$_2$S atmosphere. a}, Evolution of the radial scan measured along the $h$ axis between $h = 3.8$ and $h = 8.4$ after annealing at a temperature higher than 370$^{\circ}$C (from a to d) and subsequent stabilization at 300$^{\circ}$C (from d to g); the dashed blue profile at the bottom is measured at room temperature before the annealing and is superimposed (with no offset) to
curve a to highlight the increment of the background. \textbf{b}, Same radial scan along
the $h$ axis, before (black) and after (red) partial sulphurization leading to a Janus structure. \textbf{c}, Same radial scan along the $h$ axis, before (black) and after (red) complete sulphurization of the PtSe$_2$ layer; The black arrow indicates the expected position for a freestanding (unstrained) PtS$_2$ Bragg reflection.  \textbf{d}, In-plane reciprocal space map equivalent to that shown in the inset of Fig. \ref{fig:xrd_asgrown}a corresponding to a fully sulphurized PtS$_2$: the peak positions remain unaltered with respect to the as-grown PtSe$_2$ case, although chalcogen substitution is demonstrated by photoemission spectroscopy. Remarkably, most of the peaks display diffuse scattering tails pointing toward the Pt(440) reflection. \textbf{e}, Scans along the out-of-plane direction $l$ perpendicular to the sample surface at the positions ($33l$), ($50l$) and ($70l$), relative to a  PtSe$_2$ Bragg peak and two superstructure peaks, before (orange) and after (green) the complete sulphurization of the PtSe$_2$ monolayer. \textbf{f}, Pictorial scheme of the sulphurization process for the transformation of PtSe$_2$ into a Janus material: PtSe$_2$ is first annealed in vacuum at temperatures slightly higher than the growth temperature (370$^{\circ}$C); subsequently it is exposed to H$_2$S gas at 350$^{\circ}$C where the defects and vacancies created during the first step become ideal sites for H$_2$S adsorption and Se-by-S substitution. The process ends up with the conversion of the pristine PtSe$_2$ into a Janus SPtSe material.\\

Figue 3: \textbf{XPS measurements on PtSe$_2$, SPtSe and sulfide samples.} Se and S CLs of the different samples studied in this work, \textit{i.e.} a pristine non-sulphurized PtSe$_2$ (green), the Janus SPtSe (red), and a fully substituted PtS$_2$ (blue). The take-off angle is 45$^{\circ}$. From \textbf{a} to \textbf{c} different binding energy ranges are shown: \textbf{a}, Se \textit{3s} and S \textit{2s}, \textbf{b}, Se \textit{3p} and S \textit{2d}, and \textbf{c}, Se \textit{3d}. All the samples were first annealed in vacuum prior to the measurements.\\

Figure 4: \textbf{Angle-resolved XPS core-level spectra of Janus SPtSe. a}, Sketch illustrating the AR-XPS technique and the relationship between the probing depth $d$ and the take-off angle $\theta$ ($d=\lambda sin(\theta)$, where $\lambda$ is the attenuation length taking into account elastic scattering\cite{nefedov2}). The probing depths are 11.5 \AA, 28.2 \AA\ and 47.15 \AA\ for 10$^{\circ}$, 25$^{\circ}$ and 45$^{\circ}$ take-off angles respectively. \textbf{b-d}, X-ray photoelectron spectra of Se \textit{3p} and S \textit{2p} core levels, taken at 45$^\circ$, 25$^\circ$ and 10$^\circ$ take-off angle $\theta$ on the sulphurized sample. The red line above the spectra corresponds to the fitting residual. The spectra have been normalized to the intensity at the S \textit{2p} core level. \textbf{e}, S/Se concentration ratio as a function of the take-off angle $\theta$.

\newpage

\newpage

\newpage

\newpage

\end{document}